%% file: draft.tex
\documentclass[11pt]{article} 
\usepackage[utf8]{inputenc}
\usepackage{multirow}
\usepackage{amsfonts}
\usepackage{amsmath}
\usepackage{amssymb}
\usepackage{amsthm}
\usepackage{caption}
\usepackage[dvipsnames]{xcolor}
    \definecolor{darkgreen}{rgb}{0,0.5,0}
    \definecolor{darkblue}{rgb}{0,0,0.6}
    \definecolor{purple}{rgb}{0.4,.2,0.7}
\usepackage[margin = 2.5cm]{geometry}
\usepackage{graphicx}
\usepackage[hyperfootnotes = false, colorlinks = true, linkcolor = darkblue, citecolor = purple]{hyperref}
\usepackage{subcaption}
\usepackage{ytableau}

\usepackage[font=small,skip=3pt]{caption}
\usepackage{wrapfig}
\usepackage{lipsum}
\usepackage{mathrsfs}

\def\la{\label}
\def\nref#1{(\ref{#1})}
\def\half{{1 \over 2 }}

\def\tr{\mathrm{tr}}

\newcommand{\bh}{\hat b}
	\newcommand{\bes}{\begin{equation} \begin{split} }	
	\newcommand{\ees}{\end{split} \end{equation} }

	\newcommand{\lp}{\left (}
	\newcommand{\rp}{\right )}

	\usepackage{physics}

\newcommand{\be}{\begin{equation}}
\newcommand{\ee}{\end{equation}}
\newcommand{\bea}{\begin{eqnarray}}
\newcommand{\eea}{\end{eqnarray}}
\def\nref#1{(\ref{#1})}
\def\half{{1 \over 2 }}
\def\tr{\mathrm{tr}}

\begin{document}

\thispagestyle{empty}
\begin{center}
    ~\vspace{5mm}

  {\LARGE \bf {Comparing the decoherence effects due to black holes versus ordinary matter\\}} 
    
   \vspace{0.5in}
     
   {\bf    Anna Biggs$^1$ and  Juan Maldacena$^2$ 
   }

    \vspace{0.5in}

  $^1$
  Jadwin Hall, Princeton University,  Princeton, NJ 08540, USA 
   \\
   ~
   \\
  $^2$
  Institute for Advanced Study,  Princeton, NJ 08540, USA

    \vspace{0.5in}

    \vspace{0.5in}
    

\end{center}

\vspace{0.5in}

\begin{abstract}
Recently a certain thought experiment was discussed which involves the decoherence of a quantum system due to a black hole. 
Here we show how this phenomenon is consistent with standard ideas about quantum black holes. In other words, modeling the black hole as a quantum system at finite temperature one obtains the same answer. We demonstrate this by analyzing the problem in terms of an effective theory that can apply both for the black hole case and for an ordinary matter system, showing that the same qualitative effect is present for ordinary matter at finite temperature.

 \end{abstract}
 
\vspace{1in}

\pagebreak

\setcounter{tocdepth}{3}
{\hypersetup{linkcolor=black}\tableofcontents}

\input{Introduction.tex}

\input{EFTapproach.tex}

\input{Decoherence.tex}

\input{FDT.tex}

\input{AbsorptionandMatching.tex}

\input{Ordinarymatter.tex}

\input{PageSection.tex}

 \input{Conclusion.tex}

 \input{ThermalCorrelators.tex}

\eject
  
\bibliographystyle{apsrev4-1long}
\bibliography{GeneralBibliography.bib}
\end{document}

%% file: Introduction.tex
 \section{Introduction and Motivation}
 
  Recently, it was suggested that black holes exhibit a unique decoherence effect \cite{Danielson:2022sga, Danielson:2022tdw},  and that this could be of fundamental importance for their quantum description.

\begin{figure}
\captionsetup[subfigure]{margin={3.15cm,4cm}}
	\begin{subfigure}{.45\textwidth}
   \includegraphics[height=7cm]{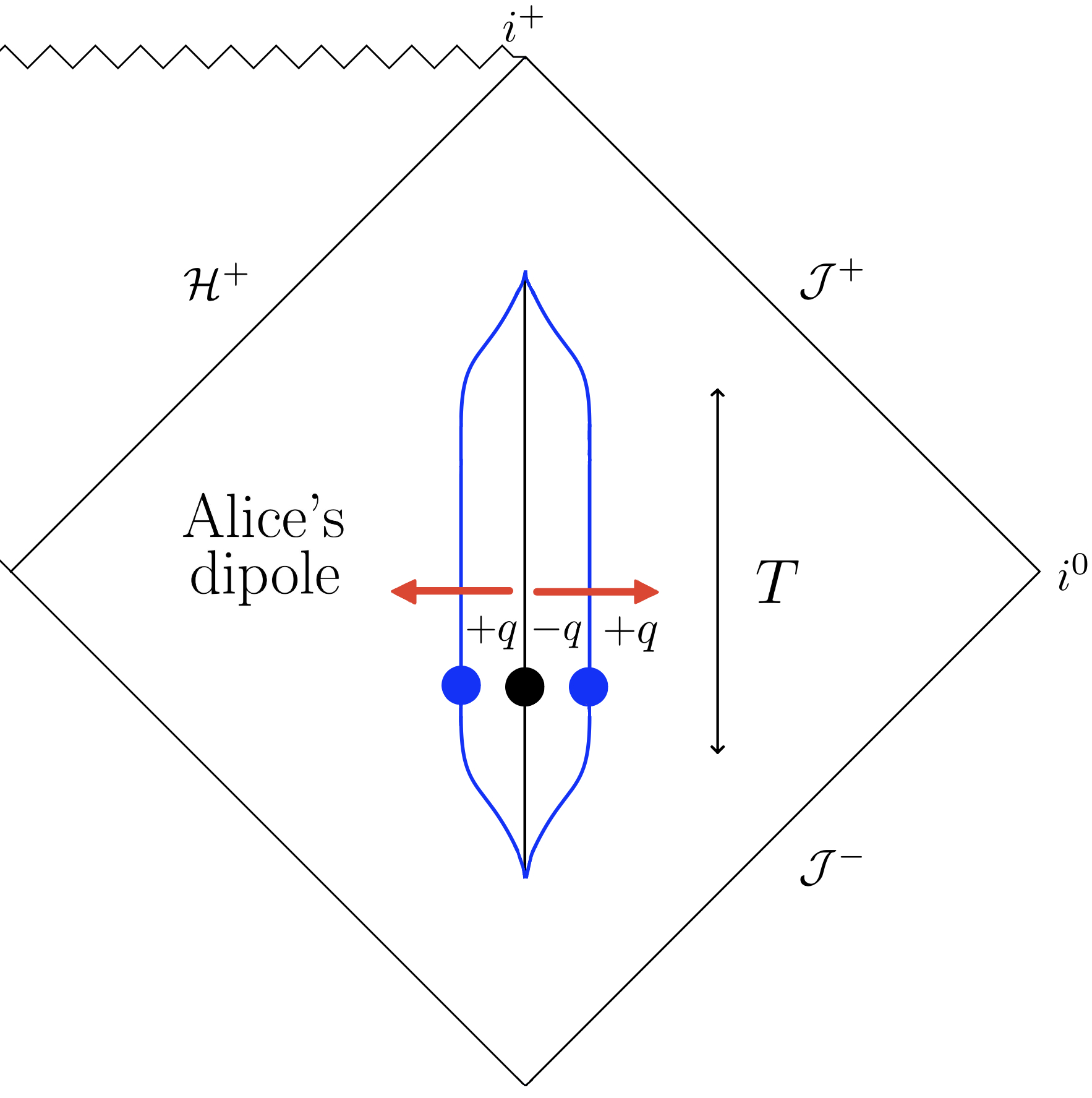}
   \caption{}
    \la{penrose}
\end{subfigure}
\captionsetup[subfigure]{margin={5cm,4cm}}
\begin{subfigure}{.5\textwidth}
 \vspace{10mm}
 \includegraphics[height=4cm]{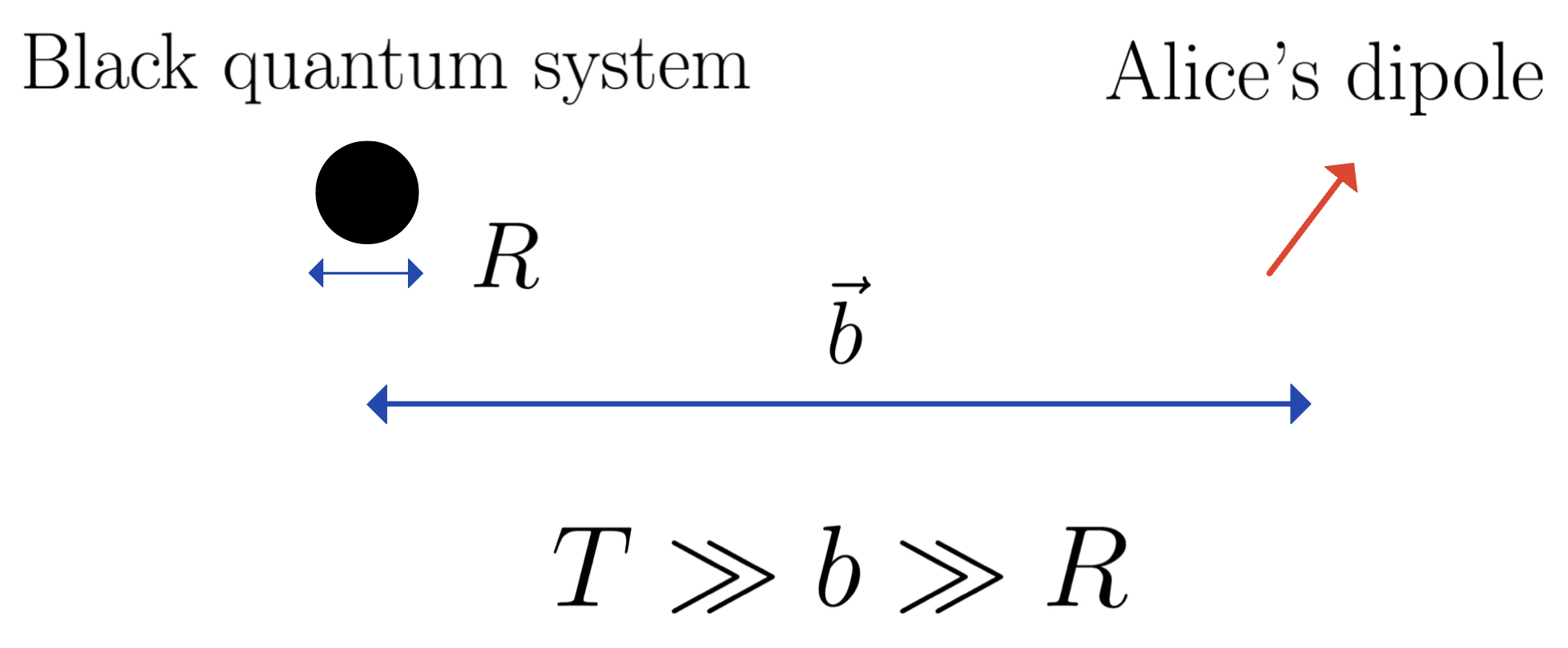}
 \vspace{12mm}
 \caption{} 
    \la{bqsalice}
\end{subfigure}
\caption{Setup for the decoherence problem discussed in \cite{Danielson:2022sga, Danielson:2022tdw, Gralla:2023oya}.  Alice prepares a spatial superposition of charged (or massive) particles in the vicinity of a black hole. The electromagnetic case is pictured here. Between the two positions of the $+q$ charge there is a static  particle of charge $-q$. So, her state is a superposition of an electric dipole pointing in two opposite directions. Alice holds the superposition stationary for a long proper time $T$, during which the presence of the horizon destroys its coherence. (a) Spacetime Penrose diagram. (b) Picture from the outside. The black hole could be replaced by a quantum system, which we call the ``black quantum system."}
\vspace{2mm}
\end{figure}



Here we explain that this decoherence phenomenon can be understood using  standard ideas about quantum black holes, namely,  the idea that we can replace the black hole by a quantum system with a number of qubits of order the black hole entropy. (This is sometimes called the ``central dogma'' of quantum black holes.)

We analyze the thought experiment proposed in \cite{Danielson:2022sga, Danielson:2022tdw, Gralla:2023oya} (see Figure \ref{penrose}) in terms of an effective theory where we replace the black hole by a quantum system. This effective theory then applies equally well for a black hole as for ordinary matter, which enables us to also compute the effect for the case of ordinary matter. The setup involves two quantum systems, one is Alice's system and the other is what we will call a ``black quantum system'' which could be a black hole or a piece of ordinary matter.

We show that the effects found in \cite{Danielson:2022sga, Danielson:2022tdw, Gralla:2023oya} can be viewed as arising due to thermal or quantum fluctuations of multipole operators  
describing the black quantum system. In fact, the decoherence effect  can be described purely in terms of two point functions of these multipole operators.  These two point functions govern other interesting observables such as the  the absorption cross sections. In addition,  the two point function is related to the dynamical response function, or dynamical Love number, of the black quantum system through the fluctuation dissipation theorem or KMS (Kubo-Martin-Schwinger) condition.

In other words, we can describe the decoherence phenomenon in terms of an effective theory which only involves operators describing the electric dipole moment or mass quadrupole moment of the black quantum system. The parameters of this effective theory can be read off from a (low frequency) computation of Schwarzschild observables such as the dynamical Love numbers or the absorption cross section. 
Note that even though the static Love numbers are zero for black holes, there is a first order in frequency, $\omega$, correction which is non-zero and is responsible for the effect in question \cite{Fang:2005qq, Damour:2009vw, Binnington:2009bb, Kol:2011vg, Chakrabarti:2013lua, Hui:2020xxx, Poisson_2004, Chia:2020yla, Charalambous:2021mea, Ivanov:2022hlo, Ivanov:2022qqt, Perry:2023wmm, Saketh:2023bul, Mariadraft}.



Comparing the black hole case with ordinary matter at the same temperature we find a qualitatively similar effect, as long at the entropy of matter is macroscopic. The effect depends on the resistivity or viscosity of the material. 
For the electromagnetic case, it is relatively easy to find quantitatively similar effects. This is not surprising because we can have black objects which absorb electromagnetic fields.   
For the gravitational case, the effect is weaker for ordinary matter, if we consider objects of the same size as the black hole. This is just a reflection of the smallness of the Newton constant, or the weakness of gravity's coupling to matter.

The organization of the rest of the paper is as follows. In section \ref{EFTapproach},  we state the effective theory  picture. In section \ref{Deco}, we derive an expression for the decoherence of Alice's superposition in terms of two point functions of the effective theory operators. In sections  \ref{FDTsec} and \ref{abssec}, we discuss the relation between the two point functions and other observables. 
In section \ref{matching}, we compute the decoherence due to a Schwarzschild black hole, reproducing previous results of \cite{Danielson:2022sga, Danielson:2022tdw, Gralla:2023oya}. In section \ref{Comp}, we estimate the decoherence due to various simple types of matter.\footnote{We work in units where $c = \hbar   = k_{B} = 1$, so that $\epsilon_{0} = 1/e^{2}$.} Finally, in section \ref{Page}, we review the implication of the recent results on the Page curve for general decoherence effects by black holes.

%% file: EFTapproach.tex
 \section{The effective theory approach}\la{EFTapproach}

\subsection{Review of the setup for the decoherence thought experiment} 

We first briefly recall the setup of the decoherence problem described in \cite{Danielson:2022sga, Danielson:2022tdw, Gralla:2023oya}. There is a stationary experimenter Alice at a radial position $b \gg R$, where $R$ is the Schwarzschild radius, or  here the size of the black quantum system, see Figure \ref{bqsalice}. She has a particle of charge $q$ which she prepares in a superposition of position eigenstates,\footnote{More realistically, we can imagine a superposition of position-space wavepackets with little overlap.} $|\Psi\rangle_{\text{Alice}} = \frac{1}{\sqrt{2}} \lp |-\frac{d}{2}\rangle + |+\frac{d}{2} \rangle \rp $ where $d$ denotes the distance between the spatial positions. 
 Alice maintains the superposition for a long proper time $T$ and then recombines the particle. The separation and recombination are performed adiabatically so that negligible radiation is emitted to future infinity. This is possible if $ T_{\text{turn}} \gg q d$, where $ T_{\text{turn}}$ is the timescale of the separation and recombination processes \cite{Belenchia_2018}. We are interested in the limit\footnote{We are not accounting for the backreaction due to Hawking radiation, and will assume that $T$ is much smaller than the lifetime of the black hole, $T \ll \tau_{\text{BH}} \sim G^{2}M^{3}$.}   where $T \gg T_{\text{turn}}$ and $T \gg b$. In \cite{Danielson:2022sga, Danielson:2022tdw} it is shown that   the black hole, and more generally any bifurcate Killing horizon, will destroy the coherence of the superposition at a rate exponential in $T$. The authors explain that this effect can be understood as a consequence of the electromagnetic fields sourced by the particle falling into the black hole. This causes decoherence because the fields in the black hole interior contain information which distinguishes $|-\frac{d}{2}\rangle$ from $|+\frac{d}{2} \rangle $. 

We make a tiny modification of the setup where we include a stationary negative charge $-q$ so that we see more clearly that the two branches of the wavefunction for Alice's system correspond to an electric dipole that is pointing in two opposite directions. See Figure \ref{penrose}. From now on we will denote Alice's dipole by $\vec P_{A}$. Alice's quantum system then effectively consists of a single qubit where the $\sigma_3$ eigenstates correspond to the two opposite directions for this dipole.

\subsection{The effective theory for the electromagnetic case}

We now state a first effective field theory which is a logical stepping stone to the effective theory that we will eventually use.  For this we note that the fields sourced by $\vec{P}_{A}$ have frequency of order $\omega \sim 1/T$. Since  $T \gg b \gg R$, their wavelength is much larger than the size of the black quantum system, see Figure \ref{bqsalice}. In this limit the black quantum system can be approximated as a point particle in flat spacetime.
The interactions of the point particle with low-frequency bulk fields are captured by multipole operators living on the point particle worldline.\footnote{This ``worldline effective theory'' is implicitly assumed in usual discussions of black holes as quantum systems and was used in string theory approaches to black hole thermodynamics. It was also discussed in  \cite{Goldberger:2004jt, Goldberger:2005cd}, see \cite{Goldberger:2022ebt} or \cite{Porto_2016} for a review. 
}
 For example, the interaction which governs the scattering of electric fields off the body takes the form
\begin{align}
	S_{\text{int}}^{e} &= -\int dt ~\vec P_{B} (t) \cdot \vec E(t) \la{Sinte}
\end{align}
and the full action then takes the form 
\be \la{AcEM}
S= S_{\rm Blackbody}  + { 1 \over 2 e^2 } \int d^{4} x(\vec E^2 - \vec B^2 ) -\int dt ~\vec P_{B}(t)\cdot \vec E(t) -\int dt ~\vec P_{A} (t)\cdot \vec E(t)  \sigma_3 
\ee 
where $S_{\rm Blackbody}$ governs the black quantum system and $\vec P_B$ is an operator acting on that system. Alice's quantum system is just a single qubit and $\sigma_3$ acts on it. $\vec P_A$ is  a classical vector indicating the size of the dipole. 
This first effective field theory is pictured in Figure \ref{eft1}.
\begin{figure}
\captionsetup[subfigure]{margin={3.6cm,4cm}}
\begin{subfigure}{.46\textwidth}
    \centering
    \includegraphics[width=0.8\linewidth]{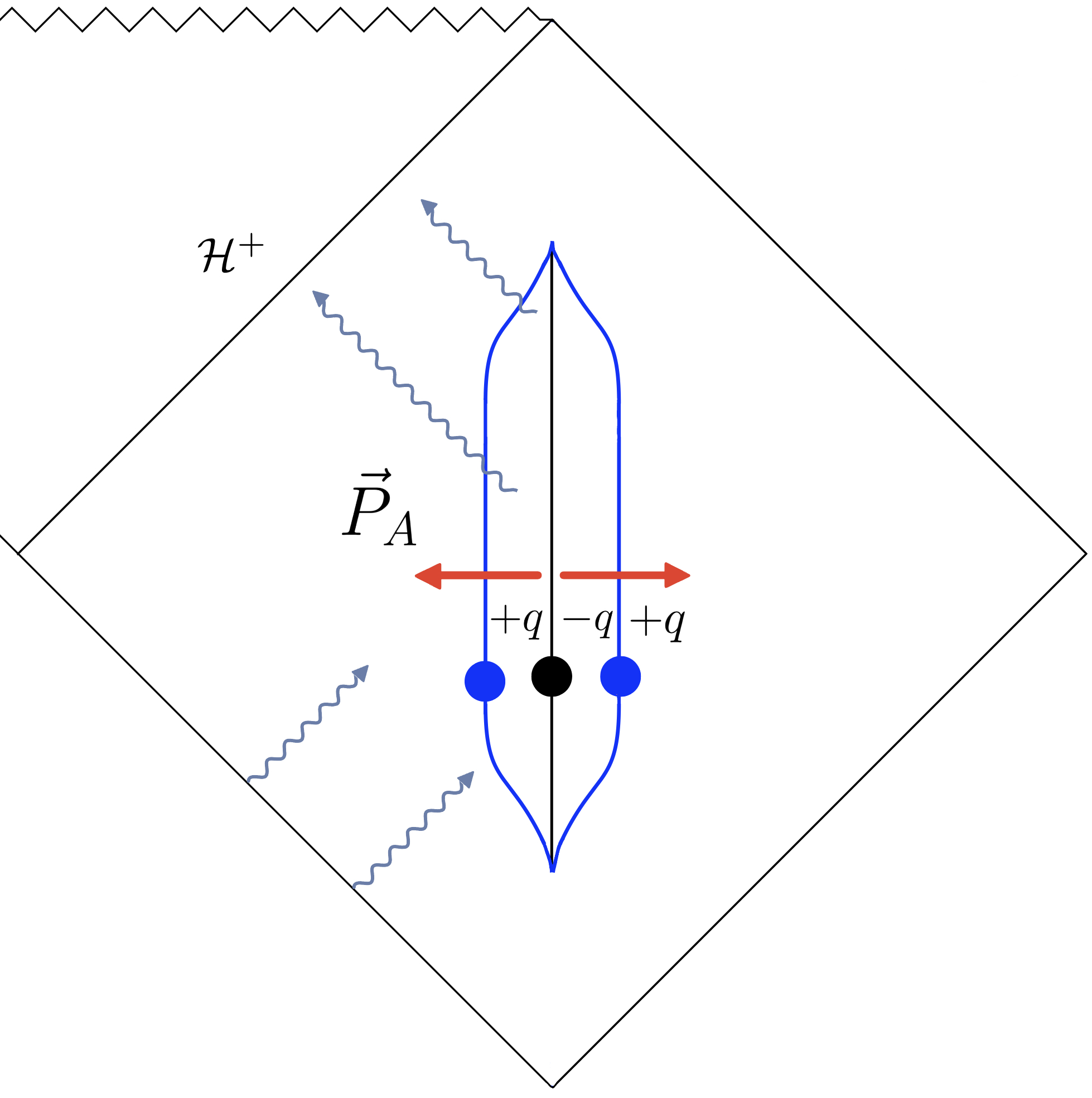}
    \caption{}
    \la{newcharge}
\end{subfigure}
\captionsetup[subfigure]{margin={1.6cm,4cm}}
\begin{subfigure}{.25\textwidth}
    \centering
    \includegraphics[width=.67\linewidth]{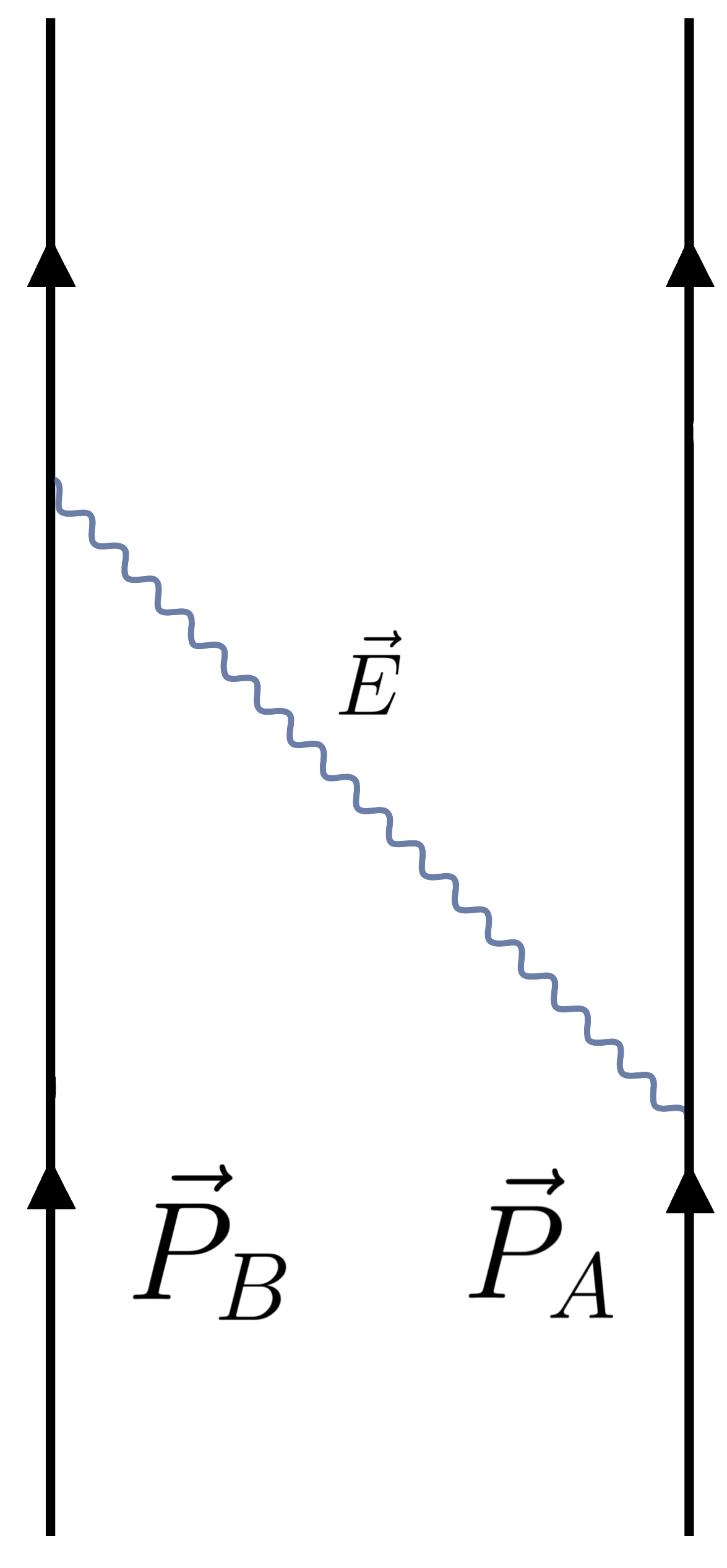}  
  ~~~  \caption{}
    \la{eft1}
\end{subfigure}
\hspace{4mm}
\captionsetup[subfigure]{margin={2cm,4cm}}
\begin{subfigure}{.25\textwidth}
    \centering
    \includegraphics[width=.67\linewidth]{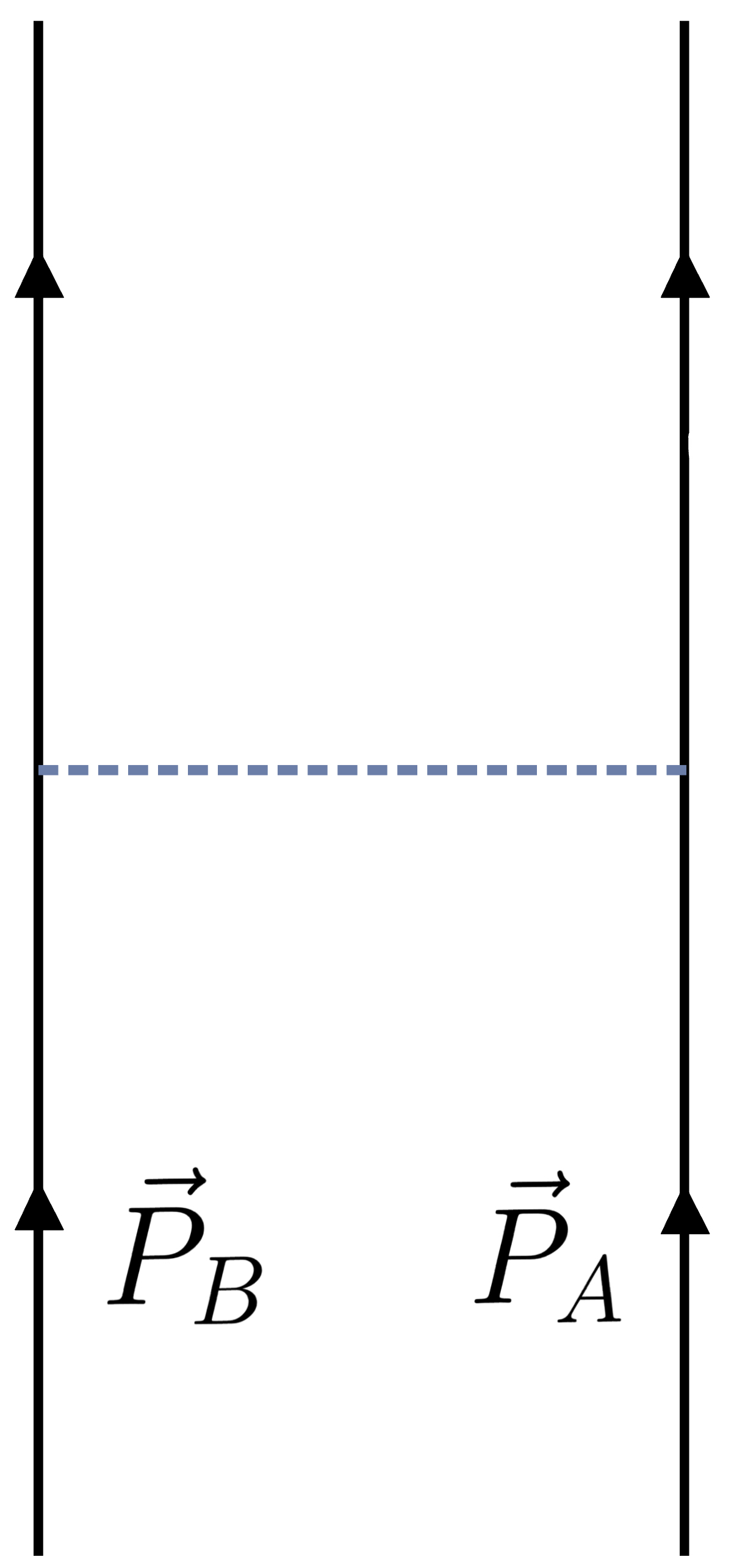}  
    \caption{}
    \la{eft2}
\end{subfigure}
\caption{Three pictures useful for understanding the decoherence effect. Alice's state is a superposition of two physical dipoles of opposite sign.
(a) We have the full spacetime picture.  (b) We replace the black hole by its electric dipole operator acting on a suitable Hilbert space. (c) We integrate out the electromagnetic field completely so that we have a direct interaction between the two dipoles.  In this approximation the interaction is instantaneous. 
}
\vspace{3mm}
\end{figure}

In general, $S_{\text{int}}^{e}$  should include all operators that respect the symmetries of the problem, which in this case includes SO(3) rotational invariance, diffeomorphism invariance, and worldline reparamaterizaiton invariance. In (\ref{Sinte}, \ref{AcEM}) we have neglected the magnetic dipole moment and all $\ell >1$ operators. Since the decoherence problem involves very slow moving sources, it is sufficient for our purposes to include only the interactions which are leading order in $R \omega$.

We are interested specifically in the interaction between the black object dipole, $\vec P_{B}$,  and the electric field sourced by Alice's dipole,  $\vec P_{A}$. So a simpler effective theory is obtained by integrating out the electric field completely, leaving only a dipole-dipole interaction. This is the effective theory we will primarily use throughout the text, and is illustrated in Figure \ref{eft2}. Integrating out $\vec{E}$ gives the effective action
\begin{align}
	S_{\text{int}}^{e} &= -\frac{e^{2}}{4 \pi b^{3}}\int dt \lp \vec{P}_{A} \cdot  \vec{P}_{B} - 3  \lp \vec{P}_{A} \cdot \hat b \rp \lp  \vec{P}_{B} \cdot \hat b \rp  \rp\sigma_3 \la{Seint}
\end{align}
where $\hat{b} = \vec{b}/|\vec{b}|$. In writing \nref{Seint} we have taken the limit where $\vec P_{A}$ varies on a timescale $T$ much larger than $b$ and much larger than the timescale of variation of $\vec P_{B}$, which is set by the inverse temperature $\beta $.

\subsection{The effective theory for the gravitational case} 

The analysis of the gravitational case is analogous. Here the leading order interaction is between the mass quadrupole moments $Q_{A}$ and $Q_{B}$  of Alice's superposition and the black quantum system, respectively. Again,  we make a slight modification of the setup to highlight that the effect involves Alice's mass quadrupole moment. 
 While we cannot add particles of negative mass,  Alice can create a superposition of two configurations with opposite mass quadrupole moment in the following way. We imagine there is a stationary mass distribution that is unchanged by the experiment. We add a stationary mass dipole by displacing some of the mass from the background mass density. Then Alice will create an additional dipole that she can move between two alternate locations on either side of the stationary dipole, see Figure \ref{quadrupole}.

\begin{figure}[h] 
\begin{center}\includegraphics[height=4.5cm]{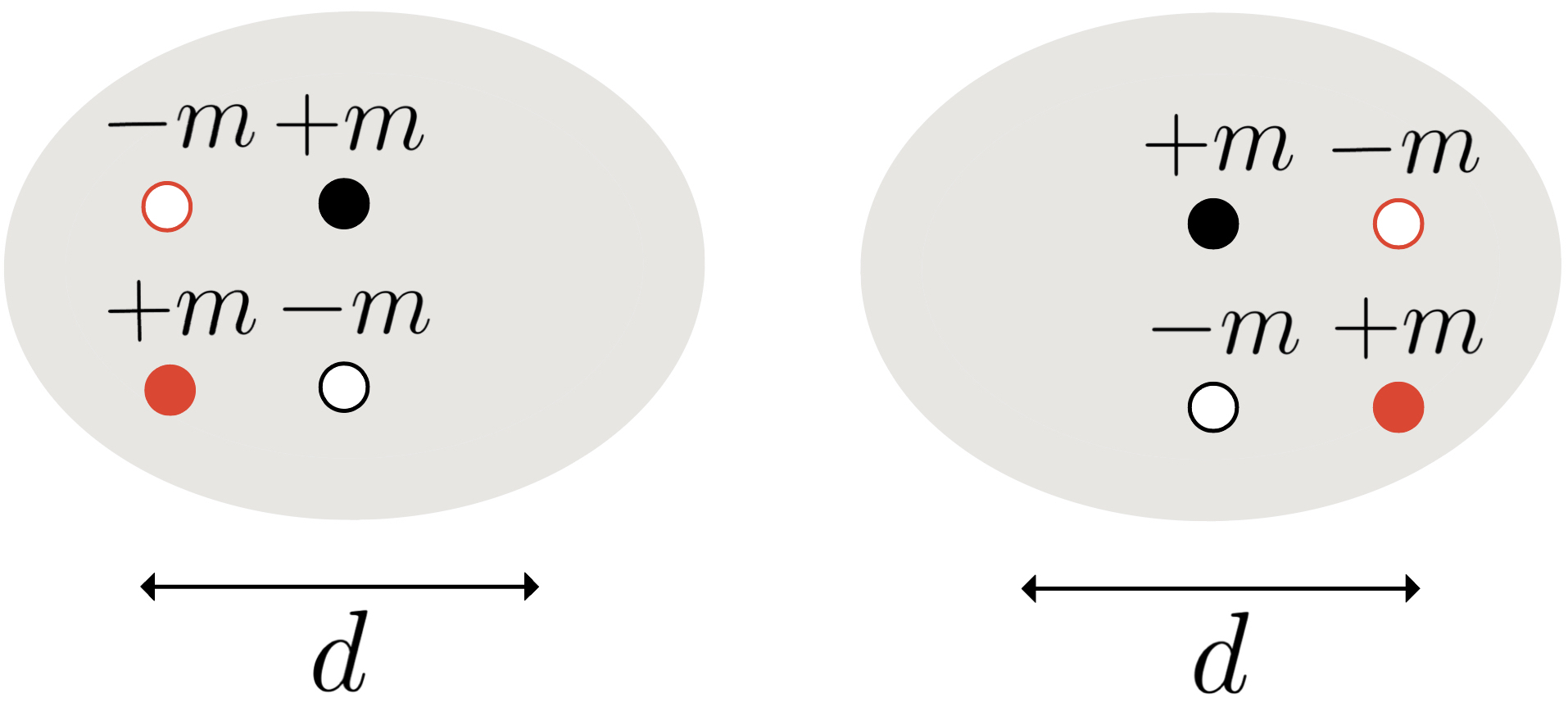}
\caption{A convenient setup for the gravitational experiment. There is a background stationary mass distribution with pockets of higher and lower density. The experiment consists of moving a (red) mass dipole to the left or to the right, with a stationary (black) dipole in the center. Alice's state is then a superposition of two quadrupoles of equal magnitude and opposite sign.  } \la{quadrupole}
\end{center}
\label{etsetup}
\end{figure}

This mass quadrupole interacts with a component of the Riemann curvature. Again we denote by $Q^{ij}_A$ the classical size of the quadrupole of Alice's system, and $\sigma_3$ is the operator that acts on the qubit degree of freedom specifying the sign of this quadrupole. 

Beginning with stress energy sources corresponding to Alice's dipole and the black quantum system and integrating out the graviton, we find that the interaction between the two mass quadrupoles in the limit where $T \gg b$ and $T \gg \beta$ is
\begin{align} 
	S_{\text{int}}^{g} &= \frac{G}{12b^{5}} \int dt \lp 2 \,\tr \lp Q^{A}  Q^{B}\rp+ 35 \lp \hat b^{\intercal }Q^{A} \hat b \rp \lp \hat b^{\intercal }  Q^{B} \hat b \rp -20 \lp \hat b^{\intercal} Q^{A} Q^{B} \hat b\rp  \rp\sigma_{3} \la{Sgint}
\end{align}
$Q_{ij}(t)$ is the traceless quadrupole moment. For an ordinary matter distribution, $Q_{ij}$  is defined as
\begin{align}
	Q_{ij}(t) = \int d^{3}x ~\rho(t,x)\lp3  x_{i}x_{j} - x^{2}\delta_{ij}\rp
\end{align}

\subsection{The effective theory for the scalar field case } 

Finally, following \cite{Gralla:2023oya}, we can also consider a decoherence effect for spin zero fields, which we imagine are sourced by a scalar charge   $j$. Unlike in the electromagnetic and gravitational cases, $j$ is not conserved. So, as pointed out in \cite{Gralla:2023oya}, we can consider an experiment where the two branches of Alice's superposition evolve with different time-dependent charge densities $ \pm j_{A}(t)$ and remain at the same spacetime location. Each will act as a monopole source for the scalar field. 

In this case we have the action 
\be 
S = -\half \int d^{4}x (\nabla \phi)^2 + \int dt j_B \phi + \int dt j_A \phi \sigma_3 
\ee 
where $j_A$ is the strength of the (non-conserved) scalar charge. 
Integrating out the scalar field we find 
\begin{align}
S_{\text{int}}^{s} &=\frac{1}{ 4 \pi b} \int dt ~ j_{A}(t) j_{B}(t)\sigma_{3}\la{Ssint}
\end{align}

\subsection{Side comment} 
  
One minor side comment is the following. In all these cases, we can think of the two possible orientations of Alice's qubit as defining two different Hamiltonians for the black quantum system. The two Hamiltonians have an extra multipole operator inserted whose sign depends on the orientation of Alice's qubit. The decoherence arises because the black   quantum  system states evolve to orthogonal  states under these two possible Hamiltonian evolutions. An interesting related effect was discussed in \cite{Almheiri:2021jwq}.

%% file: Decoherence.tex
 \section{Decoherence in terms of the two point correlation function}\la{Deco}

In this section we show how the interaction can create some decoherence. 
For this purpose we now assume that the interaction has the form 
\be 
 S_{int}  = \int dt O(t) \sigma^3  \la{EFTsi}
 \ee 
 where $\sigma^3$ acts on the Hilbert space of Alice's quantum system which is effectively a single qubit parametrizing the two possible configurations that Alice is considering. $O$ is an operator acting on the black quantum system. 
 
 We start with Alice's system in a general state described by the two-by-two density matrix  $\rho_A$. The black quantum system starts in a thermal state $\rho_B$. Then Alice's initial state at $t=0$ evolves to $\rho_A(t)$ at time $t$, with 
 \bea 
 \rho_A(t) &=&  \tr_B \left[  \mathcal{T} e^{ i \int_0^t dt' O(t') \sigma^3 } (\rho_A \otimes \rho_B )   \overline{\mathcal{T}}e^{ -i \int_0^t dt' O(t') \sigma^3 } \right] 
\cr 
 &=&  \rho_A + ( \sigma^3 \rho_A \sigma^3 - \rho_A)  \times \int_0^t  dt' \int_0^t dt''\langle O(t') O(t'') \rangle   + \cdots  \la{Evl}
\eea
where we assumed that the one point functions of the operator $O$ vanish and we expanded only up to quadratic order in $O$. 
We now use the time translation symmetry of the thermal correlators, as well as the assumption that the two point function $\langle O(t) O(0) \rangle $ decays after a time of order $\tau_{e} \ll t$ so that
\be \la{Gammadef}
\int_0^t  dt' \int_0^t dt''\langle O(t') O(t'') \rangle \sim  t \Gamma ~,~~~~~~~\Gamma = \int_{-\infty}^{\infty } dt  \langle O(t) O(0) \rangle 
\ee 
For the black hole, the decay $\tau_{e} \sim \beta$. We now pick a time  $t$   that is long compared to $\tau_{e}$ and $b$, but short compared to $1/\Gamma$.  
Making the small    $t \Gamma$  approximation we can now write \nref{Evl} as a Lindblad  equation for the density matrix of Alice's system 
\be \la{LinEq}
\dot \rho_A = \Gamma ( \sigma^3 \rho_A \sigma^3 - \rho_A) = \Lambda \rho_A \Lambda^\dagger - \half \Lambda^\dagger \Lambda \rho_A - \half \rho_A  \Lambda^\dagger \Lambda ~,~~~~~~~~\Lambda \equiv \sqrt{\Gamma} \sigma^3 
\ee 
This equation exponentially damps the off diagonal components of the density matrix leaving just the diagonal components.\footnote{The situation described here is an example of a common model for decoherence sometimes called a ``dephasing channel'' or ``phase-damping channel.''}
More precisely, the off diagonal components decay as $e^{ - 2 \Gamma t } $. 

Notice that in deriving the Lindblad equation we assumed that after each time step $t$ the black quantum system returns to thermal equilibrium. This is called the ``Markovian" approximation.   This is approximately correct when the system is sufficiently chaotic  and we are looking at a simple operator such as $O$. 
Of course, it is not precisely true for a system with finite entropy, but we expect the corrections to be negligible for times small compared to exponentials of the entropy. 

In order to make contact with \cite{Danielson:2022sga, Danielson:2022tdw, Gralla:2023oya}, we note that they start with the pure state density matrix for Alice's system
\begin{align}
	\rho_A = |\Psi \rangle \langle \Psi | ~,~~~~~~~~~~~|\Psi \rangle = \frac{1}{\sqrt{2}} \lp \big  | - d/2 \rangle + |+d/2\rangle\rp ~,~~~~~~{\rm or } ~~~ \rho_A = \half \left( \begin{array}{cc} 1 & 1 \\ 1 & 1 \end{array} \right) 
\end{align}
where we wrote it in the qubit basis appropriate for the Hamiltonian \nref{EFTsi}. $\Gamma$ is related to the quantity denoted $\mathscr{D}$ in \cite{Danielson:2022sga, Danielson:2022tdw} by 
\be 
\la{DecoWa} 
\mathscr{D} = 2 \Gamma T
\ee

  We can now write the values of $\Gamma$ for each case by considering the appropriate operator that couples to Alice's qubit degree of freedom. For the electromagnetic case, this is described by the dipole-dipole interaction \nref{Seint}. 
In evaluating $\Gamma$, we treat Alice's dipole as a constant, classical function, turned on for the time $T$.  
 Plugging the interaction \nref{Seint} into \nref{Gammadef}, we find
\begin{align}
\la{De}
		 \Gamma_e \equiv \lp \frac{e^{2}}{4 \pi b^{3}} \rp^{2}    P_{A}^{k} P_{A}^{l} N(\hat b)_{ik} N(\hat b)_{j l} S^{ij}(\omega = 0)
\end{align}
where we use the following notation for the Wightman function:
\begin{align}
	S^{ij}(\omega) \equiv \int dt e^{i \omega t}\langle P_{B}^{i}(t) P_{B}^{j} (0)\rangle
\end{align}
and $N_{ij}$ is a tensor describing the dependence on the relative orientation of the two dipoles.
\begin{align}
	N(\hat b)_{ij} = \delta_{ij} - 3 \hat b_{i}\hat b_{j}
\end{align}
We see that, in the large $T$ limit, $\Gamma$ is only sensitive to the correlator near zero frequency. More precisely, it is sensitive to the correlator at frequencies $\omega \sim \Gamma$, but we are assuming that $\Gamma$ is very small.

 A similar analysis in the gravitational and scalar cases gives
\begin{align}
	 \Gamma_{g} &=   \lp \frac{G}{12b^{5}}\rp^{2}   Q_{A}^{k l}Q_{A}^{cd}C(\hat b)_{ijkl} C(\hat b)_{abcd}S^{ij ab}(\omega = 0)  \la{Dg}\\
\Gamma_{s} &=  \lp \frac{1}{4 \pi b}\rp^{2}  j_{A}j_{A} S(\omega = 0)\la{Ds}
\end{align}
where 
\begin{align}
S^{ijkl}(\omega) &\equiv \int dt e^{i \omega t} \langle  Q^{ij}_{B}(t)  Q^{kl}_{B}(0) \rangle\\
	S(\omega) & \equiv \int dt e^{i \omega t} \langle  j_{B}(t) j_{B}(0) \rangle
	\end{align}
and $C_{ijkl}$ is the tensor
\begin{align}
	C(x)_{ijkl} = 2\delta_{ik}\delta_{jl}+35 \bh_{i} \bh_{j}\bh_{k}\bh_{l}-20 \bh_{i}\delta_{jk}\bh_{l}
\end{align}

We see that the decoherence can be expressed in terms of two point functions of dynamical multipole operators describing the black quantum system. The precise form of these two point functions depends on the system in question.

\subsection{A qualitative picture of the decoherence}

A qualitative picture of this decoherence effect is as follows. As an example let us consider the electromagnetic case; the others are similar. The interaction Hamiltonian involves the dipole moment of the black object. If this were a time-dependent classical function, then the state would acquire a phase that depends on the integral of this function. This phase is opposite for the two orientations of Alice's dipole.  Now, the dipole moment of the black object  has thermal fluctuations that are coherent on a timescale $\tau_{e}$ set by the quasinormal modes, which is of order $\beta$ for a black hole.
For longer times these fluctuations are uncorrelated. We can then view the phase as a random walk, and the leading decoherence effect involves the average of the square of this phase, or the average variance of this random walk, which increases linearly with $T$. And the amplitude of the effect involves the typical size of these fluctuations, which is set by the integral of the two point function.

%% file: FDT.tex
 \section{The Fluctuation-Dissipation Theorem}\la{FDTsec}

In principle, the two point functions $S(\omega)$ depend on the microscopic details of the black quantum system. For ordinary matter systems it would contain information about transport coefficients such as conductivity, viscosity, etc. 
To make this connection, it is useful to recall the fluctuation dissipation theorem \cite{Callen, Kubo}, which relates $S(\omega)$ to the dissipative part of the response function:
\bea\la{FDTeq}
	S(\omega) &= & 2 \lp n_{b}(\omega)+1 \rp \text{Im}~ \chi(\omega)\quad  \quad \quad n_{b}(\omega) \equiv \frac{1}{e^{\beta \omega}-1}~,~
	\cr 
	& ~& ~~~~{\rm and}~~~ n_b(\omega) \sim { 1 \over \beta \omega } ~,~~~~~~~{\rm for} ~~\omega \beta \ll1 
\eea
where we have indicated that we will be interested in frequencies smaller than the temperature of the black quantum system.  See appendix \ref{thermcorr} for the derivation. The response function 
$\chi(\omega)$,  also known as the retarded Green's function, describes  the expectation value of an operator due to a small perturbation by an external source that couples to that operator. The precise definition is given in appendix \ref{thermcorr}.   For example, when an electric field is applied to a conductor, the system acquires a dipole moment which is proportional, via $\chi(\omega)$, to the external field.
$\chi(\omega)$ is also called the ``dynamical Love number'' and we will be interested in its low energy expansion, which is of the form 
\be \la{DLN} 
\chi(\omega) = A + i \omega B  + \mathcal{O}(\omega^2 ) 
\ee 
with real $A$ and $B \geq 0$ for small $\omega$. For black holes in 4d the static response $A=0$, but $B$ is non-zero, which through \nref{FDTeq} implies that $S(\omega)$ has a finite limit as $\omega \to 0$ set by the constant $B$ in \nref{DLN}. For general matter systems, which are not black holes, $A$ is typically non-zero. 

We also see that $S(\omega)$ depends only on $\text{Im}\,\chi(\omega)$, which is also called the spectral function. The imaginary part of $\chi(\omega)$   arises from absorptive processes. This alludes to an important connection between the decoherence effect and the absorption of low-energy fields, discussed further in section \ref{matching}.

Thermalization in AdS/CFT has been studied extensively. To name just a few examples, see \cite{deBoer:2008gu, Son:2009vu, Das:2010yw, Caron-Huot:2011vtx, Giecold:2009cg, Faulkner:2010jy, Hubeny:2010ry, Caldeira:2022gfo}, and \cite{Liu:2018crr} for a review of recent developments in non-equilibrium dynamics from holography.

%% file: AbsorptionandMatching.tex
\section{The connection to absorption}\la{abssec}

In this section, we point out that the same two point correlation function involved in the decoherence discussion also governs the absorption cross section of low energy waves.

The connection between classical absorption in gravitational backgrounds and correlators of a non-gravitational quantum mechanical theory is well known. Notably, this was the idea behind the matching of D-brane greybody factors calculated from both bulk gravity and the worldvolume gauge theory which gave early evidence for AdS/CFT \cite{Das:1996wn,Klebanov:1997kc, Gubser:1997se, Gubser:1997yh}.

Let us discuss the absorption of scalar fields for simplicity. The amplitude for a transition from an initial state $|i \rangle$ to a final state $|f\rangle$ of the black quantum system in a time $T$ is  
\begin{align}
	\mathcal{M}_{i \to f} = \frac{1}{\sqrt{2 \omega}}\int_{0}^{T} dt e^{-i \omega t} \langle f | \mathcal{O}(t) | i \rangle
\end{align}
To get the total transition rate we square the amplitude and sum over final states,
\begin{align}
	\Gamma_{i \to f} = \frac{T}{2 \omega}\int_{0}^{T} dt e^{i \omega t}\langle \mathcal{O}(t) \mathcal{O}(0)\rangle
\end{align}
where we have assumed that the correlator depends only on the time difference. If there are many initial states then they would also be averaged over. For example, when the black quantum system is in a finite temperature equilibrium state, this average produces a thermal correlator.

To get the absorption cross section, we divide the transition rate by the total time and the incoming particle flux. This yields
\begin{align}\la{abs}
	\sigma_{\text{abs}}(\omega) = \frac{1}{2 \omega}\int dt e^{i \omega t} \langle \mathcal{O}(t)\mathcal{O}(0) \rangle = \frac{1}{2 \omega}S(\omega) =  (n_b +1) \frac{1}{\omega} \text{Im} \,\chi(\omega) 
	\quad \quad ~~~\omega R \ll 1
\end{align}
In the case of black holes, this absorption cross section is the one computed using the full Hartle-Hawking vacuum in the near horizon geometry. It is related to the classical absorption cross section by 
\be 
\sigma_{\text{abs}}(\omega) = (n_b+1)  \sigma_{\text{abs}}^{\text{clas} }(\omega) ~,~~~~~~~ \longrightarrow ~~~~ \sigma_{\text{abs}}^{\text{clas} }(\omega) = { {\rm Im} \chi \over \omega } \sim { \beta \over 2 } S(\omega) ~,~~~~{ \rm for } ~~\omega \beta \ll 1 \la{clasabs}
\ee

In the case of electromagnetism we have a couple of modifications. First, a low energy electromagnetic wave couples both to the electric dipole moment as well as to the magnetic one, which means that there are two dipole operators. In addition, the coupling involves the electric field, which contains a derivative that introduces an extra factor of $\omega$ in the amplitude. The cross section contains an extra factor of  $ e^{2}\omega^2$. 
However, it is still the case that the answer involves the   zero frequency part of the two point function of the electric and magnetic dipoles.  

In the gravitational case, again we have a second quadrupole operator beyond the one we have discussed. These operators couple to the Riemann curvature, which means that we have an extra factor of $\omega^2 $ in the amplitude  and final factor  $\omega^4 G_N$ in the cross section.

In conclusion, two point functions of the multipole operators describing the black quantum system measure its ability to absorb low-frequency bulk fields. Using \nref{abs} and the analogous expressions for the electric and gravitational cases, $\Gamma$ could equivalently be expressed directly in terms of the absorption cross section. This connection also allows us to make contact with the perspective of \cite{Danielson:2022sga, Danielson:2022tdw, Gralla:2023oya}, where the decoherence is understood as arising due to the absorption of low frequency fields by the black hole.

\section{Matching to black holes}\la{matching}

When the black quantum system is a black hole, we generally do not have  a microscopic description which would allow us to compute $S(\omega)$ directly.\footnote{Actually, we do have microscopic descriptions in special cases in string theory.} However, $S(\omega)$ can be obtained by matching with known black hole scattering amplitudes or dynamical Love numbers. 

The matching can be done to a variety of Schwarzschild observables. For example, we can compute the absorption cross section as in \nref{abs} using the effective theory action, then equate this with the equivalent absorption cross section in the Schwarzschild background to read off $S(\omega)$. 

Another option would be to match response functions. In the effective theory, $\chi(\omega)$ is directly related to $S(\omega)$ by the fluctuation-dissipation theorem \nref{FDTeq}. On the gravitational side, the black hole response to long wavelength external fields is described by the tidal response coefficients or so-called Love numbers. Note that while the static Love numbers vanish for 4d Schwarzschild black holes, the dissipative response or ``dynamical Love numbers"  do not, and this first correction is imaginary and linear $ i \omega$.   It is this part of the response function which contributes to $\text{Im}\chi(\omega)$. 
    Schematically, $\chi(\omega)$ is computed in gravity as follows. We solve the wave equation for the external field in Schwarzschild with ingoing boundary conditions at the horizon.  One finds  a solution to the radial equation in the region $ r_s \ll r \ll 1/\omega$ which scales as $r^{\ell}$ and $1/r^{\ell + 1}$. The former is interpreted as setting up the source field, and the latter is response of the black hole. 
$\chi(\omega)$ is proportional to the ratio between the coefficients of the $r^{\ell}$ and $1/r^{\ell + 1}$ solutions. When the external field is time-dependent, the    response is $\omega$-dependent. There is a large body of literature on dynamical black hole Love numbers. Recent papers include \cite{Perry:2023wmm, Saketh:2023bul}.

Either of these matching procedures suffice for our purposes. We will not write the details of the matching computation here; the results can be found in the literature \cite{Goldberger:2019sya, Goldberger:2020geb}. 

We now turn to checking that the decoherence rates found in \cite{Danielson:2022sga, Danielson:2022tdw, Gralla:2023oya} can be reproduced by thinking in terms of the effective theory variables described above. 

\subsection{The electromagnetic case}

In the eletromagnetic case, matching the effective theory absorption cross section to the appropriate absorption cross section in Schwarzschild  gives\footnote{In \cite{Fabbri} the classical absorption cross section was computed, which differs from the full quantum cross section, see \nref{clasabs}.} 
\cite{Goldberger:2005cd, Fabbri}
\begin{align}\la{SeFT}
	S_{ij}(\omega) = \frac{1}{3e^{2}}r_{s}^{3}\delta_{ij} \quad \quad \quad  r_{s} \omega \ll 1
\end{align}
To compare with \cite{Gralla:2023oya}, we take Alice's dipole $\vec{P}_{A}$ to point radially, in the $\hat b$ direction. The dipole has magnitude
$|\vec{P}_{A}| = q d/2$. Plugging these values into \nref{De},
\begin{align}\la{Deresult}
\Gamma_{e} =  \lp \frac{e^{2}}{4 \pi b^{3}} \rp^{2}  \times \frac{1}{3e^{2}} r_{s}^{3} \times  4 \lp \frac{q d}{2}\rp^{2}  = \frac{1}{4 \pi}\frac{1}{6 \pi}\frac{q^{2}e^{2}d^{2} r_{s}^{3}}{2b^{6}}
\end{align}
This is the scaling found in \cite{Danielson:2022tdw}. The numerical prefactor also matches the one computed in \cite{Gralla:2023oya}.\footnote{To compare with their result, note that $q_{\text{theirs}}^{2} = \frac{e^{2}q^{2}}{4 \pi}$.}

\subsection{The scalar case}
For scalars, the matching computation gives \cite{Goldberger:2019sya,  Fabbri, Page}
\begin{align}
	S(\omega) = 2 r_{s} \quad \quad \quad r_{s} \omega \ll 1
\end{align}
To make contact with the results of \cite{Gralla:2023oya}, we let $|j_{A}|$ evolve from $0$ to $Q/2$ and back, so $Q$ is the total difference in charge between the two branches of the superposition.
Plugging these values into \nref{Ds},
\begin{align}\la{Dsresult}
	\Gamma_{s} &= \lp \frac{1}{4 \pi b}\rp^{2}  \times 2 r_{s} \times \lp \frac{Q}{2}\rp^{2}  = \frac{1}{(4 \pi)^{2}}\frac{Q^{2}  r_{s}}{2b^{2}}
\end{align}
This is also the scaling found in \cite{Gralla:2023oya}.\footnote{The numerical prefactor here differs from that of \cite{Gralla:2023oya} due to the scalar action normalization.}

\subsection{The gravitational case}

For gravitons, the matching computation gives the
 $\langle Q_{B}Q_{B}\rangle$ correlator\footnote{The numerical prefactor of \nref{Sgrav} depends on the normalization of both $S_{\text{int}}^{g}$ and $Q_{ij}$, which we have chosen differently than some places elsewhere in the literature.} \cite{Goldberger:2020geb, Page},
\begin{align}\la{Sgrav}
	S_{ijkl}(\omega) = \frac{8}{45 \pi} G^{4}M^{5}\lp \delta_{ik} \delta_{jl} + \delta_{il}\delta_{jk} -\frac{2}{3}\delta_{ij}\delta_{kl} \rp \quad \quad \quad r_{s} \omega \ll 1
\end{align}
where $M$ is the black hole mass.

For the computation of $\Gamma_{g}$ we will not keep track of numerical prefactors but focus on the scaling with dimensionful quantities.
The quadrupole moment of the superposition is $ Q^{ij}_{A} \sim md^{2}  \hat b^{(i}\hat x^{j)}$ where $\hat x$ is a unit vector orthogonal to $\hat b$. Plugging this into \nref{Dg},
\begin{align}\la{Dgresult}
		\Gamma_{g} \propto \lp \frac{G}{b^{5} }\rp^{2}\times G^{4}M^{5} \times (m d^{2})^{2} = \frac{G^{6}M^{5}m^{2}d^{4}}{b^{10}}
\end{align}
This is also the scaling found in \cite{Danielson:2022sga}.

\subsection{Zero temperature black holes}

So far, we have considered systems at finite temperature. We could wonder whether the decoherence effect persists for zero temperature black holes. In \cite{Gralla:2023oya} they compute the decoherence due to an extremal Kerr black hole in the scalar and electromagnetic cases. For the scalar field, they find that the coherence of Alice's state does not decrease exponentially with time but instead has a power law dependence. Here we discuss a quick way to arrive at this conclusion from the effective theory reasoning.

At zero temperature, \nref{Gammadef} no longer holds. However, we can still use \nref{Evl}, which gives the time dependence of $\rho_{A}(t)$ to leading order in perturbation theory. This is valid for times $t$ such that the correction does not become large. 
The near horizon geometry of an extremal   black hole develops an $SL(2)$ approximate symmetry that fixes the form of the correlators.   A massless field corresponds to an operator of dimension $\Delta = 1$ so that we have 
\begin{align}
	 \int_{0}^{T} dt \int_{0}^{T} dt' \langle  j_{B}(t) j_{B}(t') \rangle &\propto  \int_{0}^{T} dt \int_{0}^{T} dt' \frac{1}{(t-t'-i \epsilon)^{2}}\\
	&\propto  \ln T ~~+ ~~ \text{endpoints}\la{endpoints}
\end{align}
The second term of \nref{endpoints} depends on the details of how we turn on and off the perturbation.\footnote{The end result has the form $\log( T/T_{\rm turn} ) $ where $T_{\rm turn}$ is the time scale over which we turn on and off  the scalar charge, assuming $b \ll T_{\rm turn} \ll T $.} So at zero temperature \nref{endpoints} has a logarithmic, rather than linear time dependence. This is the leading order contribution  in a perturbative expansion and is consistent with \cite{Gralla:2023oya}. We are also assuming that $T$ is not too large, so  that quantum gravity effects described by the Schwarzian mode \cite{Jensen:2016pah,Engelsoy:2016xyb,Maldacena:2016upp,Stanford:2017thb,Yang:2018gdb,Kitaev:2018wpr}   are not important, though it seems possible to include such effects too, if desired.

For operators with dimension $\Delta > 1$,
\begin{align}
	 \int_{0}^{T} dt \int_{0}^{T} dt' \langle  O(t) O(t') \rangle &\propto \int_{0}^{T} dt \int_{0}^{T} dt' \frac{1}{(t - t' - i \epsilon)^{2 \Delta}} \\
	&\propto \lp \frac{1}{T}\rp^{2(\Delta - 1)} + ~~~\text{endpoints}
\end{align}
where we see that $T$ dependent terms go to zero in the $T \to \infty$ limit. The remaining contribution to the decoherence depends on the details of  how we turn on an off the scalar charge of Alice's system. If these details are independent of $T$, then  we do not get any $T$ dependence at long times (again neglecting Schwarzian corrections).

%% file: Ordinarymatter.tex
 \section{Comparison to ordinary matter}\la{Comp}
  
 Now we address ordinary matter systems. We estimate the response function for various simple types of matter such as a conductor in the electromagnetic case and a self-gravitating fluid or elastic solid in the gravitational case. We will consider systems at finite temperature, since we have also mainly considered black holes at finite temperature. 
 
 For the systems we consider this also implies  a fairly large entropy, which means that the energy levels are exponentially close to each other and can absorb the low-frequency modes $\omega \sim 1/T$. It turns out that ordinary macroscopic materials have sufficiently large entropy that this is true for all practical purposes; that is, when $T$ is less than the age of the universe.
 
 As we explained previously, the response functions set the decoherence rate, so comparing the response functions of black holes versus those of ordinary matter is the same as comparing the decoherence effects.
 
 \subsection{Electric response by ordinary matter versus black holes}
  
 The dissipative response function for the black hole interacting with an external electric field is given by the dynamical love number \cite{Charalambous:2021mea, Ivanov:2022hlo},  
 or from (\ref{SeFT}) after using the fluctuation dissipation theorem, 
 \begin{align}
	\text{Im}\,\chi_{\text{BH}}^{e}(\omega) \sim \frac{1}{e^{2}}\omega r_{s}^{4}
\end{align}
 We want to compare this to the response function of an ordinary material. To make a fair comparison we consider objects of the same size and at the same temperature as the black hole. 
 
 \subsubsection{Conducting solid}
 
We begin by estimating the response function of a conductor subject to an external, time-varying electric field $\vec{E}_{\text{ext}}$. For this purpose we are interested only in rough estimates, and will not be careful about numerical prefactors. 

Let us consider a conducting sphere of radius $R$ and resistivity $\rho$. We first consider a static  $\vec{E}_{\text{ext}}$. The charges on the sphere   rearrange themselves so as to cancel the electric field inside the sphere, resulting in a net dipole moment. Let $\mathcal{Q}$ denote the order of magnitude of the charge on the top half of the sphere.  The electric field due to these charges is of order  $\vec{E}_{\text{cond}} \sim \frac{e^{2}\mathcal{Q}}{R^{2}}$, and the  dipole moment of the conductor is of order  $\vec{P} \sim \mathcal{Q} R$. Setting $\vec{E}_{\text{ext}} \sim  \vec{E}_{\text{cond}}$ gives $\chi = \lp \mathcal{Q} R \rp / \lp \mathcal{Q}/R^{2} \rp = \frac{1}{e^{2}}R^{3}$. This is the zero frequency response.  

When $\vec{E}_{\text{ext}}$ is time-dependent, the cancellation is incomplete due to the finite resistivity which  prevents charges from rearranging immediately.   We now have
\begin{align}\la{ohmslaw}
	\vec E_{\text{ext}} - \vec{E}_{\text{cond}} = \rho  \vec{J}
\end{align}
where $\vec{J}$ is the current density and $\rho $ the resistivity. $\vec{J}$ is related to the time derivative of the dipole moment by $\int d^{3} x \vec{J} =\frac{d \vec{P}}{dt} $, 
which implies $R^{3} \vec{J} \sim - i \omega \vec{P}$. Plugging this into \nref{ohmslaw},
 \begin{align}
 	&E_{\text{ext}} - \frac{e^{2}\mathcal{Q}}{R^{2}} \sim E_{\text{ext}} - \frac{e^{2}P}{R^{3}} \sim \rho \lp \frac{- i \omega P}{R^{3}} \rp \quad \Rightarrow \quad \vec{P} \sim  \frac{1}{e^{2}}R^{3} \lp \# + \frac{i \omega \rho}{e^{2}}   \rp \vec{E}_{\text{ext}}
  \end{align}
where we have assumed $\omega \rho \ll 1$. Therefore
  \begin{align}\la{chie}
  \text{Im} \, \chi_{\text{cond}}(\omega) \sim \frac{1}{e^{4}}\omega \rho R^{3}
  \end{align}
 To be comparable to a black hole, we see that $\rho/R$ should be of order $e^{2}$.
 
  In general, higher resistivity corresponds to greater absorption of low-frequency fields. So,  we should consider a relatively impure metal such as an aluminum alloy. The electrical resistivity of aluminum alloy at very low temperatures is typically on the order of $ \rho/\hbar \sim 60 ~\mu$m \cite{Yin2016TheMO}.\footnote{The resistivity of metals goes to a constant at low temperatures due to the scattering with impurities, which is temperature independent \cite{Chambers}. 
  Note that while pure aluminum is a Type I superconductor, aluminum alloys are not.} 

 As an example consider a black hole of size $R=r_s=50 ~\mu $m, chosen so that its temperature, $T \sim 4 $ K is not too  low. Then an aluminum ball of this size would then have 
 $\rho/R \sim 1>e^{2}$. 

In order for the conductor to absorb very low frequency modes,   its energy level spacings should be sufficiently small. For a system with sufficient interactions this is ensured if it has a  large entropy. 
A metal can be  approximated by an ideal Fermi gas. The entropy of a Fermi gas at low temperatures scales as $S  \propto \frac{N  }{\beta E_{F}}$ where $N$ is the number of particles and $E_{F}$ is the Fermi energy. For the aluminum ball, we find $S  \sim 10^{13}$ and so  $\Delta E \sim e^{-10^{13}}E_{0}$, which is exponentially small. Here $E_0$ is a characteristic energy, whose precise size does not matter for this argument, given the exponential prefactor. 

\subsubsection{Aside on the membrane paradigm}

It is well known from the membrane paradigm \cite{Thorne:1987bsa} that we can think of the black hole horizon as having a surface resistivity $\rho_s/\hbar = e^2$, or $\rho_s = 377~ \Omega$ \cite{Damour}.\footnote{ 
We are using the standard particle physics normalization of the Lagrangian. Then,  $\rho_s = 4\pi$ in \cite{Damour} becomes $\rho_s = e^2$ for us, with the same value in Ohms, of course!} This is a value achievable by choosing a suitable material, as we essentially discussed above, except that above we were using a solid ball rather than a  spherical shell. For a spherical shell, instead of \nref{chie} we get $\text{Im} \chi_{\text{cond}} \sim \frac{1}{e^{4}}\omega \rho_{s} R^{4}$.

\subsection{Gravitational response by ordinary matter versus black holes}

We now compare how soft gravitons are absorbed by ordinary matter versus black holes. The black hole response function for gravitational fields is \cite{Chakrabarti:2013lua, Poisson_2004, Chia:2020yla, Charalambous:2021mea, Ivanov:2022hlo, Perry:2023wmm, Saketh:2023bul} 
\begin{align} \la{BHRes}
	\text{Im} \, \chi_{\text{BH}}^{e}(\omega) \propto \omega \frac{r_{s}^{6}}{G}
\end{align}

We will again compare the black hole to ordinary matter at the same temperature. Since this is a gravitational interaction, we will compare the black hole both to a matter configuration with the same mass, and separately to one with the same size. Not surprisingly, we will find that an ordinary object of the same mass as a black hole absorbs much more than a black hole, while the opposite is true for an object of the same radius as a black hole.

\subsubsection{A self-gravitating fluid with the same mass as a black hole}

First we consider the case of equal mass. A simple type of matter is a self-gravitating fluid. We take a fluid of radius $R$ and mass $M$, and estimate the response of its gravitational quadrupole moment to an external gravitational potential $\Phi_{\text{ext}}$. 

Suppose first that $\Phi_{\text{ext}}$ is static.  $\Phi_{\text{ext}}$ corresponds to a background curvature $\mathcal{R} \sim \partial^{2}\Phi_{\text{ext}}$, which will induce a change in the mass distribution  of the fluid, $\delta m$,  which in turn generates its own potential, $\delta \phi \sim G \delta m/R$. The mass redistribution will be such that $\delta \phi = \Phi_{\text{ext}}$. Solving for $\delta m$, $\delta m \sim R^{3}\mathcal{R}/G$. The quadrupole moment $Q$ generated by the mass deformation is
\begin{align}
	Q_{ij} \sim \delta m R^{2} \sim \frac{R^{5}}{G}\mathcal{R}
\end{align}
So the static response is $\chi(\omega = 0) \sim R^{5}/G$. Now let $\Phi_{\text{ext}}$ oscillate with a small frequency,  $\omega R \ll 1$. The nonzero total potential generates a matter current, 
\begin{align}\la{massflow}
	{ d \delta   m  \over d t} \sim \rho v R^{2}
\end{align}
where $v$ is the velocity of a fluid element and $\rho$ is the mass density. The matter flow is related to the potential by the Navier-Stokes equation,
\begin{align}\la{NS}
	\nu \nabla^{2} v \sim \nabla \lp \Phi_{\text{ext}} - \delta \phi \rp
\end{align}
where we have dropped terms that are second order in $\omega$. Here $\nu$ is the kinematic viscosity. 
Solving for $v$ from \nref{massflow}, plugging this into \nref{NS}, and expanding in $\omega \nu R/M \ll 1$,
\begin{align}\la{resp}
\chi(\omega) &\sim \frac{R^{5}}{G} \lp \#+ i \omega \frac{\nu R}{G M} \rp \quad \quad~~~~ \text{Im} \,\chi(\omega) \sim  \omega \frac{\nu R^{6}}{G R_{s}}
\end{align}
where we have used $\rho R^{3} = M$. Here $R_{s}$ denotes the Schwarzschild radius of a black hole with the same total mass as the fluid. 

It is interesting to note that if we interpret the black hole as a viscous fluid, we find 
\begin{align}\la{nuBH}
	\nu_{\text{BH}} \sim  r_{s} c
\end{align}
where we momentarily restore $c$ to make the following point. From kinetic theory, the shear viscosity $\eta$ of a fluid can be estimated as
\begin{align}
	\eta \sim \rho ~ \ell_{\text{mfp}}\langle v \rangle
\end{align}
where $\ell_{\text{mfp}}$ is the mean free path of a particle and $\langle v \rangle $
is the average velocity of a particle. So the largest possible $\nu$ would roughly be when $\ell_{\text{mfp}} = r$ and $\langle v \rangle = c$, which is the kinematic viscosity of a black hole \nref{nuBH} above.\footnote{Thanks to Zihan Zhou for pointing this out to us.} This indicates that, when interpreted as a fluid, black holes have a very high kinematic viscosity relative to other types of matter.\footnote{This is contrast with the $\eta/s =1/(4\pi)$ ratio for black holes \cite{Kovtun:2004de}, which is very small compared to ordinary matter.} 

Then the ratio of the responses for an sphere of fluid with the same mass as a black hole is 
\begin{align}\la{comp}
	\frac{\text{Im} \chi_{\text{fluid}}}{\text{Im} \chi_{\text{BH}}^{g}} \sim \frac{\nu_{\rm fluid}}{\nu_{\text{BH}}} \lp \frac{R}{r_{s}}\rp^{6}  
	\end{align}

We consider a self-gravitating ball of water at room temperature.\footnote{We are ignoring the fact that water boils at zero pressure and room temperature. We   imagine surrounding the ball with a thin film preventing its evaporation. 
} A black hole at this temperature has a radius $r_{s} \sim 0.6 ~\mu $m and the same mass as a sphere of water of radius $R \sim  5 \times 10^5 \text{ m}$.
 Using that the kinematic viscosity of water at room temperature is 
$\nu \sim 10^{-6} ~ \text{m}^{2}/\text{s}$ we find 
\begin{align}
	\frac{R}{r_{s}}  \sim 10^{12} ~~~~~~~~~~~~~~~~~~~~~~ \frac{\nu}{\nu_{\text{BH}}} \sim 10^{-8} ~~~~~~~~~~~~~~~~~~~~\frac{\text{Im} \chi_{\text{fluid}}}{\text{Im} \chi_{\text{BH}}^{g}} \sim \frac{\nu}{\nu_{\text{BH}}} \lp \frac{R}{r_{s}} \rp^{6} \sim 10^{64}
\end{align}
which implies that the ball of water absorbs gravitons more easily than a black hole of the same mass. 
	
Again, since we have a macroscopic amount of water, the entropy is large enough to ensure that the level spacings are small enough  for all practical purposes. 

\subsubsection{An elastic solid of the same size as a black hole}

Since most liquids freeze at the low temperatures corresponding to macroscopic black holes, we consider an elastic solid with some viscosity. 

 As before, the gravitational potential induces a background curvature $\mathcal{R} \sim \partial_{x}^{2}\Phi_{\text{ext}}$, which in turn causes a deformation $\vec{u}$ of the material body,
\begin{align}\la{elas1}
	\partial_{t}^{2} \vec{u} \sim - \nabla \Phi_{\text{ext}}\sim x \mathcal{R}
\end{align}
The displacement $\vec{u}$ is governed by wave equations propagating transverse and longitudinal modes.
 For the purpose of this estimate we will not distinguish between the transverse and longitudinal sound waves and simply write $c_{s}$ for the speed of sound. Similarly, we will not distinguish between shear and bulk viscosity.    
 The wave equation is then  
\begin{align}\la{elasticwe}
	\partial_{t}^{2}\vec{u} \sim  \lp c_{s}^{2} -i \omega \nu  \rp \nabla^{2}\vec{u}
\end{align} 
where  $\nu$ includes the effects of  the kinematic viscosities and thermal conductivity. Equating \nref{elas1} and \nref{elasticwe} and expanding in $\omega \nu/c_{s}^{2} \ll 1$, we find the quadrupole moment
\begin{align}
	Q \sim \int d^{3} x \rho(x) x u \sim R^{3} \rho R^{4}  \frac{1}{c_{s}^{2}}\lp \# + i \omega \frac{\nu}{c_{s}^{2}}\rp\mathcal{R}
\end{align}
So, the dissipative response function which describes the quadrupole moment tidally induced by $\Phi_{\text{ext}}$ is
\begin{align}
\text{Im} \chi (\omega)_{\text{metal}} \sim \omega \nu R^{3} \rho R^{4} \frac{c^{2}}{c_{s}^{4}} \sim \omega \nu \frac{R^{5}}{G}\lp \frac{R_{s}}{R}\rp \lp \frac{c}{c_{s}}\rp^{4}\la{chimetal}
\end{align}
where we have restored $c$ and rewritten $R_{s} \sim \rho R^{3} G/c^{2}$ for the Schwarzschild radius of a black hole with the same mass as the metal ball. Note that this is not the Schwarzschild radius of the black hole we are comparing to, which is $r_s=R$. 
Using  \nref{BHRes} and \nref{nuBH},  the 
ratio of response functions is 
\begin{align}
	\frac{\text{Im} \chi_{\text{metal}}}{\text{Im} \chi_{\text{BH}}} \sim \frac{\nu}{\nu_{\text{BH}}}\lp \frac{R_{s}}{R}\rp \lp \frac{c}{c_{s}}\rp^{4}
\end{align}
While the ratio $c/c_{s}$ is large, the Schwarzschild radius, $R_s$, of a black hole of the same mass as the metal is  much smaller than its physical size. This is a consequence of the weakness of gravity, since the Schwarzschild radius is suppressed by a factor of $G$. We already see that $\chi_{\text{metal}}$ contains an extra factor of $G$ relative to $\chi_{\text{BH}}$ by comparing \nref{chimetal} and \nref{BHRes}. 

As an example, we consider a black hole of size about $100 ~\mu$m, which has a temperature of   $\sim 1.5$ K. 
We compare with a similar sized lead ball. At this temperature,  $\nu$ for lead is $\nu \sim  10^{-5} ~\text{m}^2/\text{s}$ and the speeds of sound are of order 
$c_{\ell} \sim c_{t} \sim 10^{-6}c  $
\cite{Mason}. The Schwarzschild radius  of a black hole with the same mass as the ball of lead, $R_s$, is of order of the Planck scale. Then we find
\begin{align}
	\frac{\nu}{\nu_{\text{BH}}} \sim 10^{-10} ~~~~~~~~~~~~\frac{R_{s}}{R} \sim 10^{-30} ~~~~~~~~~\lp\frac{c}{c_{s}}\rp^{4} \sim 10^{21}
	~,~~~~\to ~~~~~~~~\frac{\text{Im} \chi_{\text{metal}}}{\text{Im} \chi_{\text{BH}}^{g}} \sim 10^{-19}
\end{align}
In this case, we see that the metal does absorb fewer gravitons than a black hole, or would produce a smaller decoherence effect, though the same qualitative effect is present.

%% file: PageSection.tex
\section{More general decoherence effects and the Page curve} 
\la{Page}

 The idea that black holes cause a unique form of decoherence goes back to Hawking 
\cite{Hawking:1976ra}.    

Similar comparisons to the ones we have performed in this paper were discussed about 30 years ago in the context of the connection between D-branes and black holes, see e.g. \cite{Das:1996wn, Gubser:1997se}.  In fact, more was understood in those cases, since the explicit quantum mechanical system that describes the black hole is known in some special cases.  

All we have done here is to rephrase that discussion in the context of ordinary Schwarzschild black holes, for which the same paradigm is expected to work, though we do not know the explicit quantum mechanical dual. 

Very interesting recent developments  \cite{Penington:2019npb,Almheiri:2019psf} (see \cite{Almheiri:2020cfm} for a review) can also be rephrased in terms of bounds for   the decoherence effects of black holes. 

Namely, Alice could have a large number of qubits as opposed to a single qubit.   In fact, it could be a number that is larger than the black hole entropy. If Alice starts with all these qubits in a low entropy state, then the presence of the black hole will decohere them and increase the entropy of her system. 
The leading semiclassical   solution would lead one to believe that the entropy of Alice's system can increase to a value larger than the initial black hole entropy, provided that the Hilbert space of Alice's system has  a large enough dimension.  
However, as Page argued in \cite{Page:1993wv}, the picture of a black hole as an ordinary quantum system implies that the entropy of Alice's system cannot become larger than the entropy of the initial black hole.

In this discussion we are assuming that Alice's system does not introduce extra energy into the black hole (but this case can also be treated). The black hole might evaporate, or we might have a mirror that reflects the Hawking radiation back to the black hole, after interacting with Alice's system. 

The recent work has shown that the entropy of Alice's system computed using the modern fine grained gravitational entropy formulas of Ryu-Takayanagi-$\cdots$-Engelhardt-Wall gives a result for Alice's system that never exceeds the entropy of the original black hole \cite{Penington:2019npb,Almheiri:2019psf}. Actually, since Alice's system is in a gravitating region the proper formulas to use were conjectured in \cite{Bousso:2022hlz,Bousso:2023sya}.

%% file: Conclusion.tex
 \section{Conclusion}
 
 In this paper, we studied the decoherence due to black holes using an effective theory that is equally applicable to both black holes and ordinary matter systems.  We showed that the same qualitative effect is present for ordinary matter at finite temperature. In other words, the decoherence effect is consistent with the hypothesis that, from the outside, black holes are described by ordinary quantum systems.  
In this framework, the decoherence is seen to arise from thermal fluctuations of the multipole moments of the black hole/matter system. For the electromagnetic effect, the decoherence can be of equal magnitude for black holes and ordinary objects. For the gravitational effect, ordinary matter produces a much weaker effect than black holes of the same size.  
 
 We also reviewed  the recent results on the Page curve which bound the amount of entropy that can be generated by the interaction of a black hole with a very complex system in Alice's  possesion. 
  
 {\bf Note:} As we were about to submit this paper,  \cite{Wilson-Gerow:2024ljx} appeared which has some overlap with the discussion in this paper. 
   
 \vspace{3mm}

\large{\textbf{Acknowledgements}}
\vspace{3mm}

\normalsize{We would like to thank Daine Danielson, Samuel Gralla,  Gerardo Ortiz, Maria Jose Rodriguez, Gautam Satishchandran, Douglas Stanford, Andy Strominger,  Bob Wald, and Zihan Zhou for helpful discussions.}

%% file: ThermalCorrelators.tex
 \appendix
 
 \section{Quick review of some properties of thermal correlators} \la{thermcorr}
 
 The response function is computed by the correlator 
 \be \la{DefGret}
 \chi(t) \equiv G_R(t) = \theta(t) i \langle [ O(t), O(0) ] \rangle~,~~~~~~~~~\chi(\omega) = \int dt e^{  i \omega t }\chi(t) 
 \ee 
 This comes from the fact that under a perturbation 
 \be 
 e^{ i \int dt g(t) O(t) } 
 \ee 
 the first order change in the expectation value of $O(t)$ is 
 \be 
 \langle O(t) \rangle = \int_{-\infty}^0 dt' i \langle [ O(t) , O(t') ] \rangle g(t') = \int dt' \chi(t-t') g(t') ~,~~~~{\rm or } ~~~~~~\langle O(\omega) \rangle = \chi(\omega) g(\omega) 
 \ee 
 Note that \nref{DefGret} implies that $\chi(\omega)$ is analytic in the upper half complex $\omega$ plane. In addition, we have 
 \be 
 \la{CondGR} \chi(\omega)^* = \chi(-\omega) ~,~~~~~~{\rm for} ~~~ \omega \in {\rm Reals} 
 \ee  for real frequencies. This implies the reality of $A$ and $B$ in the low energy expansion \nref{DLN}.  
 
  The fluctuation dissipation theorem is obtained as follows. We start by writing 
 \be 
 2 {\rm Im} \chi(\omega) = \int dt  e^{ i \omega t } \langle [ O(t) , O(0) ] \rangle 
 \ee 
 We shift the integration contour of the term containing $\langle O(0) O(t)\rangle$ to $t \to t + i \beta$ and we use the Kubo-Martin-Schwinger condition     $\langle O(0) O(t+ i \beta )\rangle=\langle O(t) O(0) \rangle $.
 Here we used that the correlators are computed on the thermal vacuum.  This then gives 
 \be \la{FDT}
  2 {\rm Im} \chi(\omega) = (1 - e^{ - \beta \omega } ) \int dt e^{ i \omega t } \langle O(t) O(0) \rangle 
  \ee 
  which relates the two types of correlators. We used that the correlators were on the thermal vacuum when we used the Kubo-Martin-Schwinger condition.

$\Im \chi(\omega)$ is related to the energy absorbed by the system. If we consider a classical source $g(t)$ and the Lagrangian $ g(t) O(t)$ then the change in energy is due to the explicit time dependence of $g(t)$,
\be 
 { d   H  \over d t } = - { d g \over dt }   O(t)  
 \ee 
 So that 
  \be \la{Chha}
 \Delta \langle H \rangle = \int_{-\infty}^\infty dt { d H \over d t }   = - \int_{-\infty}^\infty  dt dt'{ d g(t) \over dt } \chi(t-t') g(t') =  2 \int_0^\infty  { d\omega \over 2 \pi }\omega  {\rm Im}[\chi(\omega) ] |g(\omega)|^2  
  \ee 
  where we used that $g(t)$ is real so that $g(-\omega) = g(\omega)^*$ as well as \nref{CondGR}.

   We see from \nref{Chha} that  the system is absorbing energy since Im$\chi(\omega)$ is positive for $\omega > 0$. One way to see this is by inserting a complete basis of energy eigenstates and rewriting $\Im \chi(\omega)$ as
 \begin{align}
	2\Im \chi(\omega) =   \sum_{m,n}|\langle m | O |n \rangle|^{2}\lp e^{-E_{m} \beta} - e^{-E_{n}\beta}\rp 2\pi \delta(\omega-\lp E_{n} - E_{m}) \rp 
\end{align}
which is manifestly positive for $\omega > 0$.